\documentclass{aa}  
\usepackage{graphicx}
\usepackage{txfonts}
\begin{document}

   \title{SDSS low z quasar companion galaxies}



   \author{D. Bettoni
          \inst{1}
          \and
          R. Falomo\inst{1}
          \and
          S. Paiano\inst{2}
          }

   \institute {INAF - Osservatorio Astronomico di Padova Vicolo dell'Osservatorio 5, I-35122 Padova (PD), Italy\\
        \email{daniela.bettoni@inaf.it,renato.falomo@inaf.it}
         \and   
           INAF - IASF Palermo via Ugo La Malfa 153,I-90146, Palermo Italy\\
             \email{simona.paiano@inaf.it}
             }

   \date{Received XXX, ; accepted XXX}

 
  \abstract
{
We investigate the relationship between quasars and close companion galaxies using photometry and spectroscopy from the large data set of the SDSS DR16 survey.
From the SDSS-QSO catalog of quasars, we selected objects with 0.1~$<$~z~$\leq$~0.35 and absolute magnitude M(r)~$<$-21.3. For all these targets, we searched for candidate companion galaxies that have a projected distance from the target $<$~700~kpc and a radial velocity difference from the QSO $\Delta$V~$<$~1000~km/sec.
We find that in 447 QSO at least one companion galaxy is associated. 
A total of 691 associated galaxies are found in these QSO fields. In the majority of them  there is just one associated galaxy but in few cases, many companions are also discovered and in two cases the QSO is found in  rich galaxy environments.
The possible contamination due to a chance projection of the companion galaxies is found less than 5\%.
Based on the available data  we expect on average to find $\sim$2 associated companion galaxies for each QSO.
A small fraction (13\%) of the companion galaxies exhibits [OII] emission lines as signature of recent star formation. 
 However a similar fraction (16\%) of unassociated galaxies  in the same QSO fields show [OII] emission.
This study suggests that there is no significant link between the presence of these close companion galaxies or the signature of recent star formation and the QSO nuclear activity.
}


   \maketitle
%

\section{Introduction}

The impact of the environment in triggering the QSO activity is a crucial issue to understand the mechanisms of the nuclear activity \citep{Shanks_88,Shen_2017,Sanchez_02}. In spite of several studies  the interplay between local and large-scale environments and their collective influence on quasar activity is a still a debated subject (see e.g. \cite{Hewett_92}). Is it triggered by mergers, as a richer than average  galaxy environment around quasars should suggest, or not?


The nuclear activity in quasars is often assumed to occur due to a major merger of two gas-rich galaxies that feed the central engine and enable the growth of a stellar spheroid. However, details on what triggers the gas fueling and how nuclear activity affects the subsequent evolution of the host galaxies remain not fully understood. The correlations observed between the black hole (BH) mass ($M_{BH}$) and the properties of the associated galaxy bulge as the $M_{BH}$-$\sigma_{stars}$ \citep{Ferrarese_2000} or $M_{BH}$-$M_{bulge}^{*}$ \citep{Kormendy_2013} seem to point to a co-evolution of BH and spheroids but both relations did not yet clarify the role played by the environment, which is a fundamental component in investigating issues of quasar activity and its role in the evolution of galaxies. Minor and major mergers may have a key role for triggering and fueling the nuclear activity but, although  mergers are relevant as possible drivers of activity, it is still a matter of debate how this affects the AGN. The global properties of the galaxy environment are probably the main driver of the AGN activity \cite[i.e.][]{Kauffmann_2000,DiMatteo_2005}.

The study of the connection between environment and AGN activity in the nearby universe ($z<$~0.5), in recent years, has seen a great improvement thanks to the use of SDSS data that allow  for the study of large dataset of quasars. 
Using SDSS data the environment of nearby quasars was analyzed by various authors with somewhat different results. \cite{Coldwell_2006} found that nearby quasars avoid high galaxy density regions, and that the
galaxies close to quasars usually have a disc-type morphology and a high star formation rate. Later on, \cite{Strand_08} comparing the environment of nearby QSO with those of  low luminosity AGN chosen from the photometric galaxy sample of the SDSS found that quasars lie in higher density regions than the AGN at scales of less than 2 Mpc.


In the past years in a series of papers \citep{Falomo_2014,Bettoni_2015,Bettoni_2017,Stone_2021,Bettoni_23} we exploited an imaging study of both the host  galaxy properties and the environment of a sample of low redshift QSO ($z<$~0.5) in the SDSS Stripe82 area. 
With regard to the environment, \citet{Karhunen_2014} found that quasars are on average found associated with small groups of galaxies. The over-densities of galaxies are mainly observed in the closest ($<$~200~kpc) region around the source and almost vanish beyond a distance of 1 Mpc. While the imaging studies of the galaxy environments can be done for extensive sample of targets the spectroscopic study of the galaxies in the close environment  of QSO remains the unique approach  to probe the QSO-galaxy  association and thus to investigate its role for the nuclear activity.

A number of pioneering works \cite[see e.g.][and references therein]{Hutchings_2001} secured spectroscopy of galaxies close to quasars and revealed that in some cases they are at the same redshift of the QSO. Moreover many of these cases exhibit some signature of interaction from a disturbed morphology. These observations suggested that interactions between the QSO host  galaxy and the close companion might trigger and fuel the nuclear activity \citep{Araujo_23,Storchi_18,Steffen_23}. 

In a recent attempt to investigate this issue spectroscopically in a systematic way, we studied 34 low z QSO by obtaining long slit spectra of their companion galaxies (see \citep{Bettoni_2017,Stone_2021}) and 10 low $z$ QSO companion galaxies using MOS spectra \citep{Bettoni_23}.
These objects were selected from a sample of 416 quasars in the SDSS Stripe 82 area for which both the host galaxy and the large scale environments were previously investigated \citep{Falomo_2014,Karhunen_2014,Bettoni_2015}.
It was found that in about half of the targets the observed companion was associated to the QSO and in most of them some signature of recent star formation inferred from the presence of the [OII] emission line. However, the SFR is rather modest ($<$ 5 M yr$^{-1}$) and favors a scenario where the link between SF and nuclear activity is of little significance. 
For these 44 QSOs, we found that for about half (19 ) there is at least one associated companion. 
We also found that many of the associated companions and some host galaxies exhibit episodes of (recent) star formation possibly induced by past interactions \citep{Stone_2021}. However, the star formation rate (SFR) of the companion galaxies is modest, and the role of the quasar remained uncertain.

In this work, we extend the analysis of the QSO-galaxy associations for the SDSS-QSO catalogue \citep{Lyke_20} and using a much larger dataset than previous studies. 
We utilize all the available photometric and spectroscopic data from SDSS-DR16 \citep{Ahumada_2020}.
This paper is organized as follows. The data sample is presented in Section 2 and the analysis in Section 3. In section 4 are presented the main results and in section 5 the conclusions.
The results are obtained in the framework of the concordance cosmology, using $ H_{0}$ = 70 km s$^{-1}$ Mpc$^{-1}$, $ \Omega_{m} $= 0.3, $ \Omega_{\lambda}$ = 0.7. 


\section{The SDSS sample of QSO-galaxy association}

We used the SDSS DR16 {\it only-quasar} catalogue \citep{Lyke_20} to extract the targets with 0.1$<$z$\leq$0.35. This search resulted in 6700  QSO.
The latter set  contains objects classified as quasar of absolute magnitude in the range -18$<$ M(r) $<-25$. Indeed this catalogue includes very low luminosity sources that are classified as quasars.
In order to be consistent with previous studies on quasar environments  \cite[see e.g.][]{Falomo_2014} we decided to cut the set removing all targets with absolute magnitude M(r) $<$-21.3 This results in a dataset of 4544 objects.
For this QSO sample, we searched in the SDSS DR16 database \citep{Ahumada_2020} for possible spectroscopic companion galaxies in the QSO fields with a search radius of 2.5 arcmin corresponding to a maximum projected distance from the QSO in the range  $\sim$ 250 to 750 kpc.

We search for all objects classified as galaxies by SDSS that are in the redshift range 0.1 $<$ z $<$ 0.35 and with apparent magnitude $m_r<$21.0 to exclude  bad  quality spectra.
This sample is composed of 2879 galaxies in 1900 QSO fields. Of these galaxies there are 773 that have a velocity difference (based on SDSS redshift) with respect to the QSO $\Delta$V$<$ 1000 km/sec. These galaxies are distributed in 524 QSO fields. This implies that a significant fraction (27\%) of the explored QSO fields have at least 1 associated companion galaxy.


In spite of having similar redshift both the QSO and the companion galaxy, it could be that in some cases they are not really associated because of a chance coincidence of their redshifts. Although the effect is expected to be small, it is useful to get an estimate.
To this aim, we performed a test by a random assignment of a redshift to each galaxy taken from the whole set of redshift of the 2879 galaxies. Then we look again for the QSO-galaxy pairs with $\Delta$V$<$~1000~km/sec.

We performed the above test for 1000 iterations and found that the average number of false pairs  from the whole set of 2879 is 80~$\pm$~8.5, therefore  we estimate that the level of contamination by fake pairs in our sample of 2879 galaxies is $\sim$~3\%.


\begin{figure}
\includegraphics[width=8.0cm]{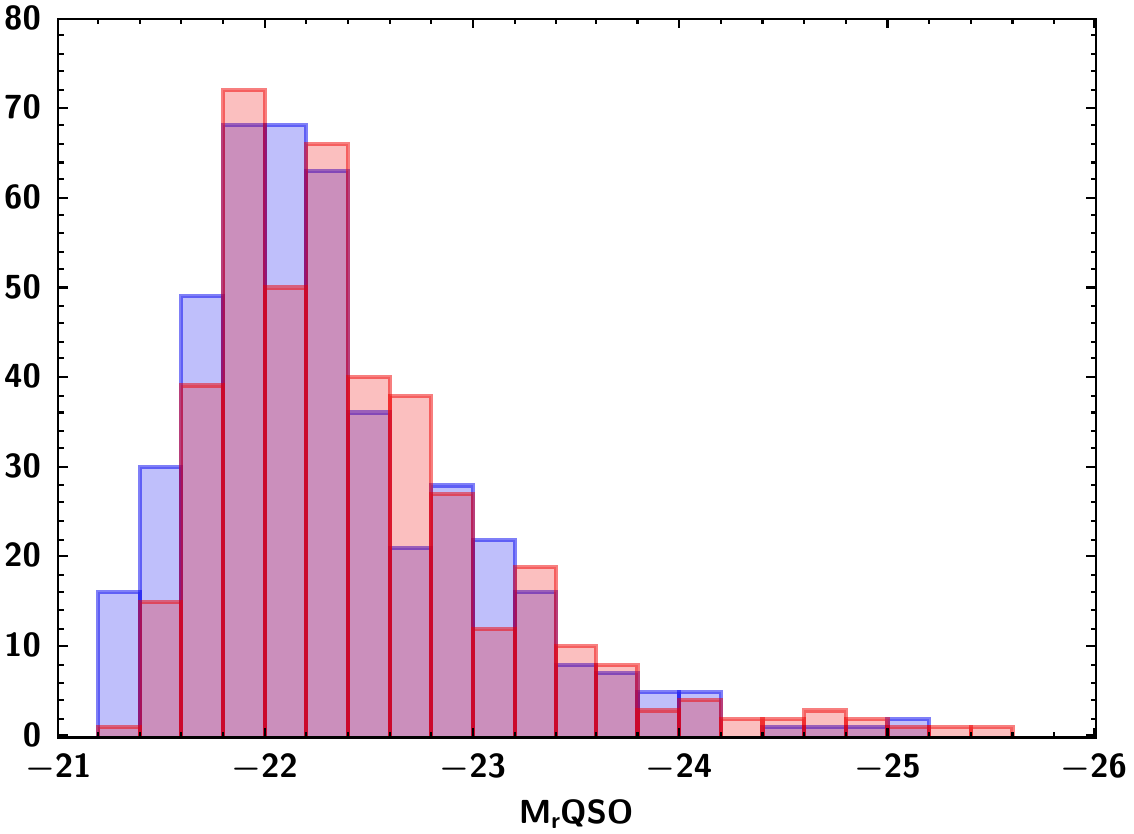}
\caption{ Absolute magnitude distribution of 447 (blue) QSO (this work) compared with the distribution of a sample of 416 QSO (red) from SDSS Stripe 82 survey \citep{Falomo_2014}.}
\label{fig:Mr_QSO}
\end{figure}

\section{Analysis of the spectra of QSO and companion galaxies}

We retrieved all the spectra both for the QSO and for the companion galaxies from SDSS DR16 database. We then measured the redshift and the equivalent width (EW) for all the emission lines of the QSOs. Since we aim to find QSO companion galaxies with $\Delta~V~<$~1000~km/s, it is important that the QSO spectrum be of adequate quality in order to allow a good measurement of the system radial velocity from the narrow emission lines. 
We therefore remove from the original list of 524 QSO those for which the spectrum was not good enough to provide an adequate quality measurement of the radial velocity.
For this purpose we check the EW of [OIII] and $H_{\beta}$ and removed from the sample all the QSO for which EW~$<$~5~\AA ~for both [OIII] and $H_{\beta}$  since below this value of EW the S/N ratio is $<$~3. 

After these revisions, the total number of usable QSO becomes 476.  
For this sub-sample, we found 691 companion galaxy candidates. In order to preserve a sound measurement of the radial velocity difference between QSO and companion galaxy, we also check the quality of the spectra and removed all galaxies with S/N~$<$~3 which implies an EW of CaII H and K absorption feature $<$~3~\AA. In addition we measured the radial velocity (RV) of both H and K and removed the objects for which the difference of RV (between the H and K lines) is $>$~500~km/s that is significantly greater than the average difference ($\sim$~200~km/s) for the whole dataset of galaxies. In these cases it turned out that at least one of the two absorption lines was very noisy and therefore their measurement is more uncertain. Thus the final sample of companion galaxies reduces to 651 sources in 447 QSO fields. This sample is very similar (see Fig.\ref{fig:Mr_QSO}) in terms of QSO luminosity to the one used in our previous study \citep{Falomo_2014} of low z QSO.

We then compared our redshift measurements of companion galaxies based on the barycenter of CaII H and K absorption lines with those provided by SDSS database that are based on the use of all features.
 We derive the difference between the redshift given by SDSS and that obtained by our measurements of the CaII H and K and MgI 5175~\AA \ absorption lines for all the companion galaxies. 
For  most of the targets the redshift difference is $<$~200~km/sec. However, in a number of cases we found a redshift difference as large as 500~km/s. 
This difference could be due to the number used features as the redshift derived by SDSS analysis is based all available spectral features that might have different quality. We carefully checked in these cases the goodness of our measurements that are based on the fit of the barycenter of H and K and a S/N~$>$~5.

Since our key parameter for the characterization of the QSO-galaxy association is the difference of radial velocity in the following we will consider only the values as derived from our measurements. 

\begin{figure*}[ht]
\hspace{-3cm}
\includegraphics[width=24cm]{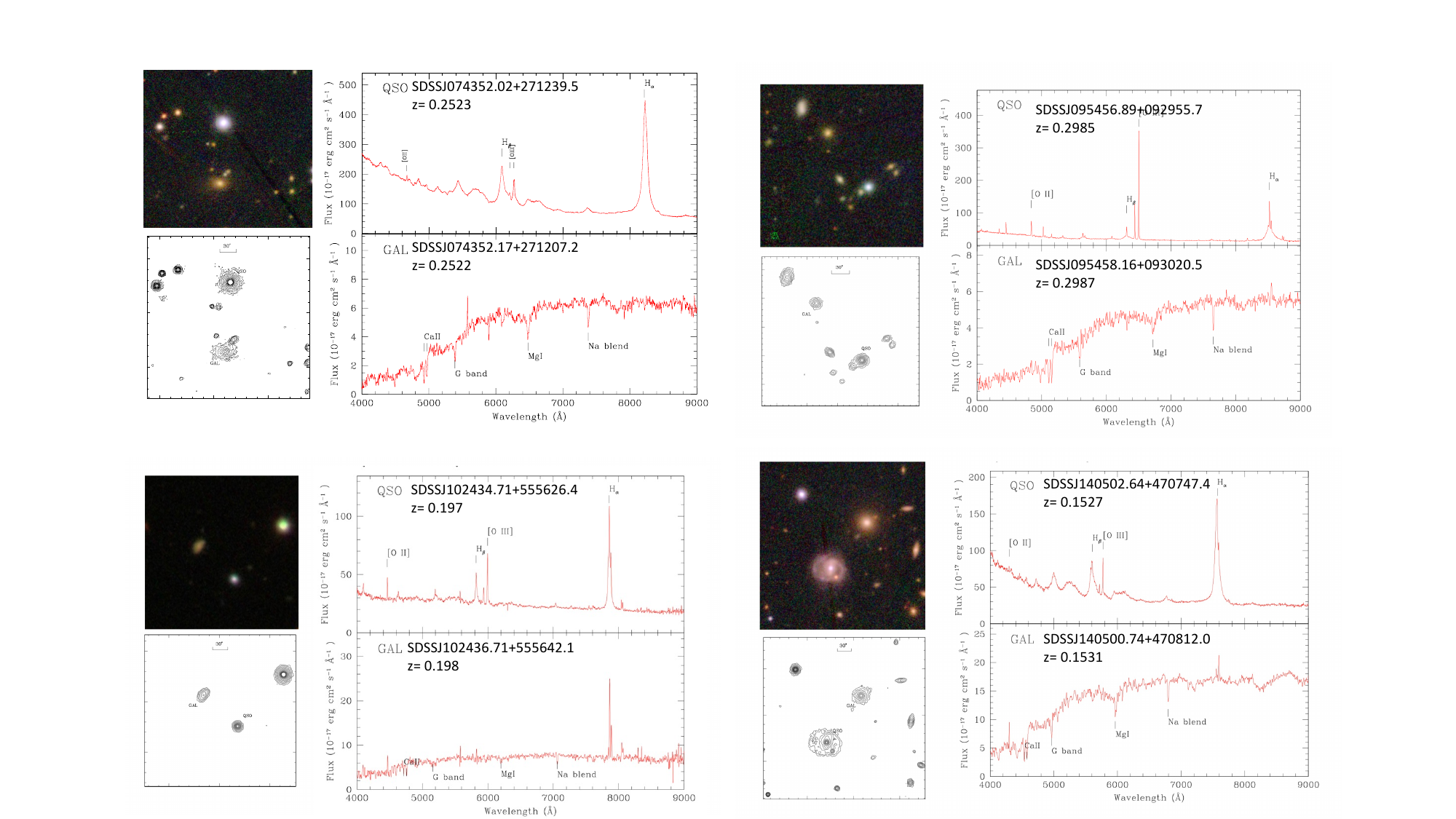}
\caption{{\it Top-Left panel}
Color image of the field  from SDSS DR16 (Field 2.5$\times$2.5 arcmin; North up East left).{\it Bottom-Left panel} Contour plot of the field. {\it Right panel} The optical spectrum of the QSO (top) and that of the companion galaxy (bottom). The main spectral features in the QSO and the companion galaxy are marked.
}
\label{fig_spec_all}
\end{figure*}

In Figure \ref{fig_spec_all} 
we report the optical spectra of four examples of QSO with their companion galaxies together with the images of the fields.

%
\section{Results}

After the cleaning of the sample (see details in Sect. 3 ), we found that in 447 QSO fields there is at least one companion galaxy with radial velocity difference $\Delta V~<$~1000~km/s. In the majority (342;  65\% ) of these fields  there is just one associated galaxy and for the remaining fields  we found 2 companions in 60 fields, and in 24 fields 3 companions, finally in 21 fields we found more than 3 companions.
In two cases we found that the QSO is in a rich group of galaxies with 8 and 12 galaxies satisfying our association criterion (see details in Appendix~A).

\begin{figure}
\includegraphics[width=9cm]{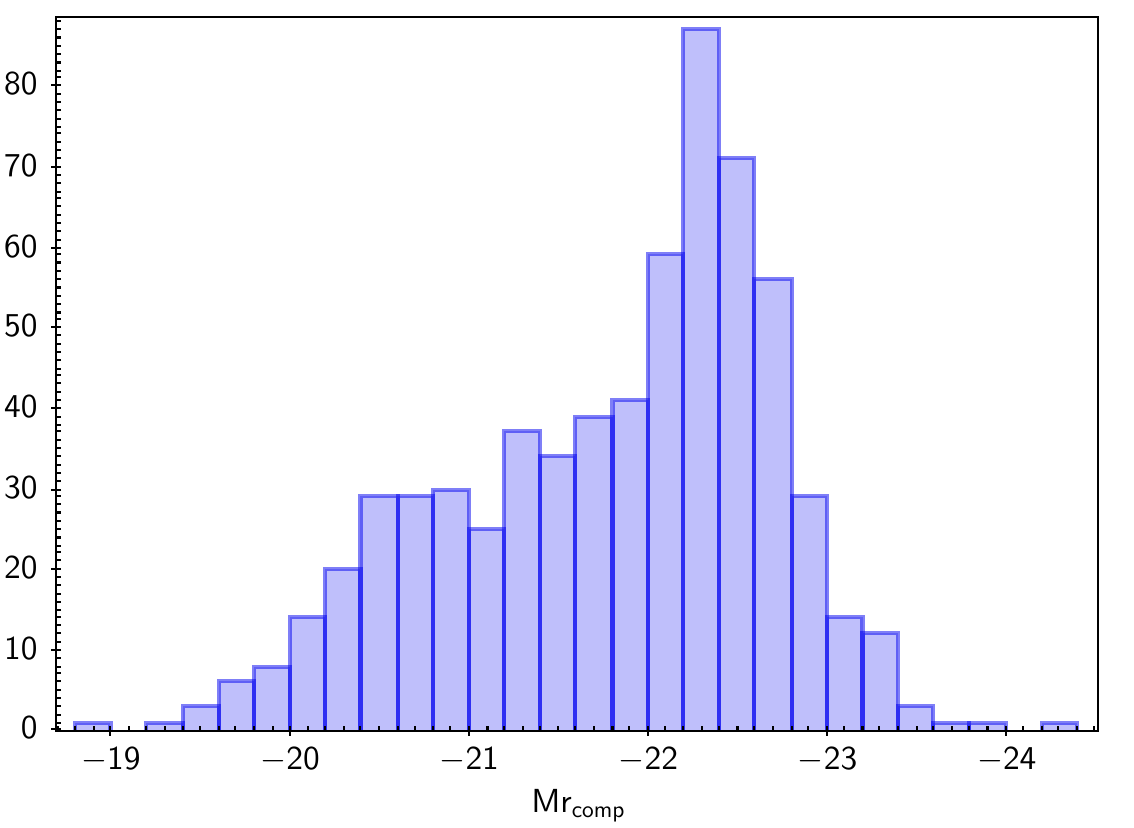}
\caption{ Absolute magnitude (M$_r$) distribution of 651 companion galaxies in 447 QSO fields (see text). The average absolute magnitude is $<M_r>$ =  --21.8 $\pm 0.9$}
\label{fig_Mr651}
\end{figure}

The distribution of the absolute magnitude (M$_r$) of the companion galaxies is shown in Figure \ref{fig_Mr651}. The companion galaxy absolute magnitude spans a wide range of values from -19.5 to -23.5 and are distributed uniformly with respect to the projected distance.
The median value of the projected distance of companion galaxies of different absolute magnitude is very similar (327~$\pm$~22~kpc, assuming four bins of absolute magnitude).
It is worth  noting that a number (32) of these companion galaxies are very luminous ($M_r <$~-23.0). The majority of these luminous galaxies are passive galaxies at $z >$~0.25, two are LINERs, and one is a Star Forming (SF) galaxy (SDSS J152551.99+181254.0) with a disk-like morphology. 
Note also that half of these luminous companion galaxies are found in QSO fields with more than one companion galaxy compared to a smaller fraction (25\%) for the companions in the whole QSO sample of 447 (see also Table 1).

%

\subsection{Statistics of companion galaxies}

Since our search for companion galaxy candidates is limited to a beam of 2.5~arcmin of radius (see Sect. 2) the explored range of projected distance of these galaxies with respect to the QSO depends on the redshift of the QSO. In order to evaluate the fraction of QSO that exhibit associated companion galaxies, we define two sub-samples that cover homogeneously the explored range of redshift and of projected distance of the companion galaxies from the QSO. These sub-samples are extracted from the whole sample described in Section 2.

\begin{figure}
\begin{center}
\includegraphics[width=9cm]{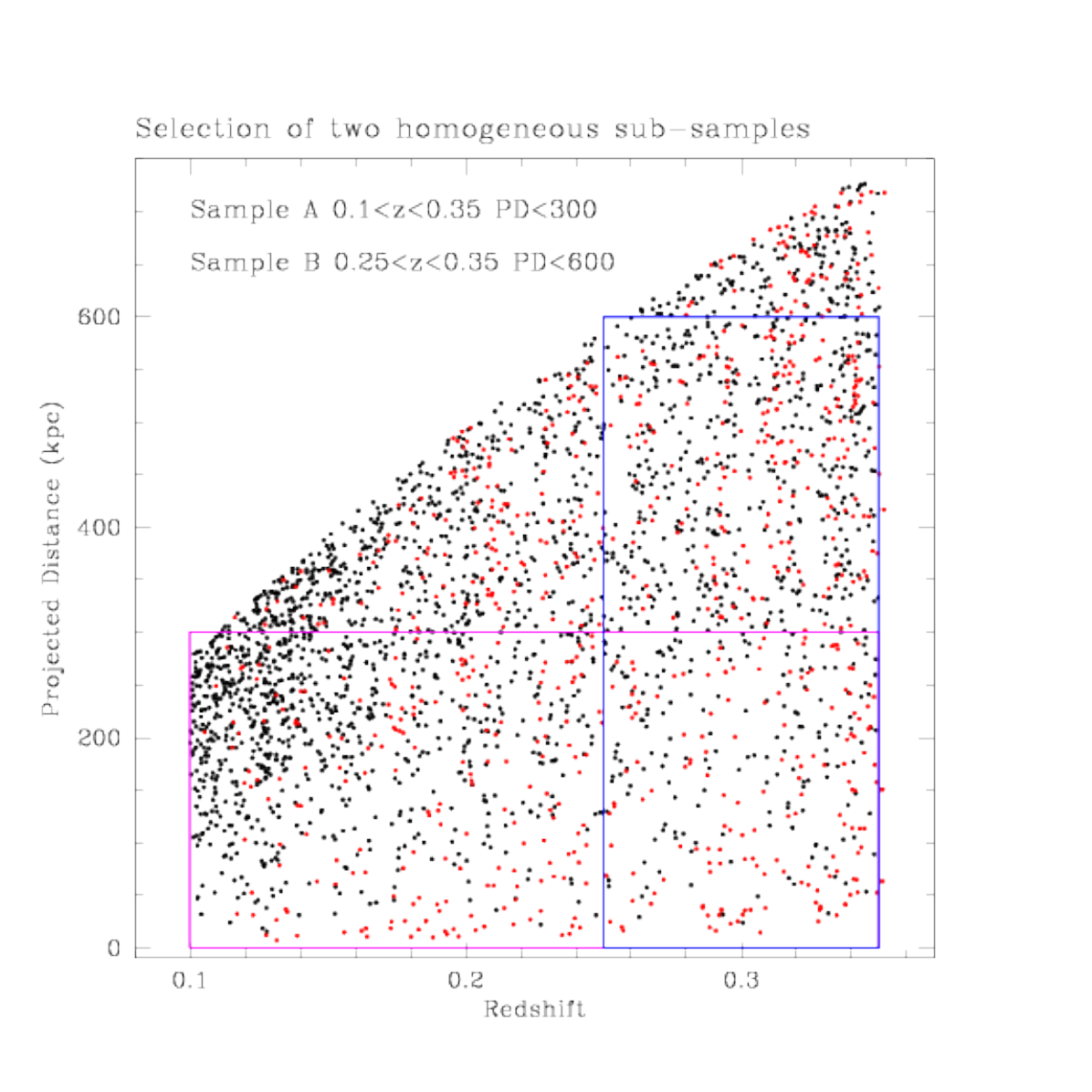}
\caption{Definition of the two subsamples of companion galaxies (see text). Red and black point represent galaxies that are associated and non associated to the QSO, respectively. 
The two rectangles define the sub-samples (see also text).
Subsample A (magenta rectangle): 0.1$<z<$0.35 and PD $<$300 kpc. 
Subsample B (blue rectangle): 0.25$<z<$0.35 and PD $<$600 kpc. 
}
\label{fig_zpd}
\end{center}
\end{figure}

The first sub-sample, sub-sample~A, was set at the maximum projected distance (PD) at 300~kpc and covers the whole range of redshift of the QSO sample (see Fig.~4). There are 1324 galaxies for 1053 QSO fields. About 30\% (355 galaxies) of them appears associated with the 297 QSO according to the $\Delta V <$~1000~km/s criterion. The other (969 objects) are non-associated galaxies  ($\Delta V >$~1000~km/s). Adopting the same test for the chance projection objects described in Sect.~2, it is found that that only $\sim$2\% of them can be spurious.

The second sub-sample,  sub-sample~B, is defined in the redshift range 0.25~$<z<$~0.35 and at a maximum PD of 600~kpc (see Fig.~4 ). There are 989 galaxies for 716 QSO fields.
About 37\% (371 galaxies) of them appears associated with the 263 QSO according to the $\Delta V <$~1000~km/s criterion. The other (618 galaxies) are non-associated ($\Delta V >$~1000~km/s). In this case the fraction of chance projection objects is of $\sim$5\%.

The statistics of the companion galaxies as compared to the non-associated galaxies in the field are summarized in Table \ref{tab_sta}.

\begin{table}
\caption{Statistics of companion galaxies for the two sub-samples of low z QSO (see text and Figure \ref{fig_zpd} )}          
\label{tab_sta}      
\centering                          
\begin{tabular}{c c c c c c}   
\hline\hline                 
Sample & $N_{g}$ & $N_Q$  & $N_c$ & $N_{Q-c}$ & N$_{Q-c}/N_Q$   \\ 
& (1) & (2) & (3) & (4) & (5)  \\
\hline \\
A & 1324  & 1053         & 355    & 297    & 0.28  \\
B &  989  & 716          & 371    & 263    & 0.37  \\
\hline                        
\hline                           
\end{tabular}
\begin{list}{}{}
\item[] Col. 1 Total number of galaxies  with spectra. Col. 2 Total number of QSO fields. Col. 3 Number of associated companion galaxies. Col 4. Number of QSO fields with associated galaxies. Col. 5 Fraction of QSO fields with associated companion galaxies. 
\end{list}
\end{table}

\subsection{Expected number of QSO companion galaxies}

In order to evaluate the frequency of expected companion galaxies in the range of mag 15.5 to 21 and with PD~$<$~700~kpc from the QSO, we  counted the number of galaxies in each field of the 4544 QSO. We found 98105 galaxies with these properties. 
However, only for 3112 QSO fields there is at least one galaxy with spectroscopic observations. Considering only these fields, we have 72042 galaxies  that satisfy the above conditions of magnitude and projected distance assuming the redshift of  the QSO. Of these companion galaxy candidates only 6379 (about 9\%) have at least one optical spectrum.

From the comparison of the redshifts of the 6379 galaxies with that of the QSO, we found that 781 galaxies (about 12\%) in 527 QSO fields the velocity difference with respect to the QSO is $<$~1000~km/s, indicating that they are likely associated with the QSO.

 Therefore in our study we found that about 17\% of the QSO have at least a companion galaxy as defined above, and for each QSO on average there are $\sim$1.5 spectroscopic companions. 
Since the spectroscopic coverage of QSO fields is limited to  $\sim$9\% of all galaxies in the fields, we would expect to find much more associated companion galaxies in case of a complete spectroscopic coverage of the galaxies in the QSO fields. 

To estimate the expected number of associated companion galaxies, we evaluated the number of associated companion galaxies (those that have $\Delta$V $<$~1000~km/s from the QSO redshift) with respect to the number of galaxies in the field that have a spectrum. This fraction reaches a peak of about 30\% around magnitude 19 and decreases at brighter and fainter magnitude of the galaxies.
Note that DR16 release includes data from other surveys (as BOSS, EBOSS, SEQUELS \citep{Dawson_2013,Dawson_16} ) with respect to standard SDSS survey \citep{Strauss_02}.
Therefore a larger number of galaxies can be used as fainter spectroscopic targets.

\begin{figure}[h]
\includegraphics[width=9.0cm]{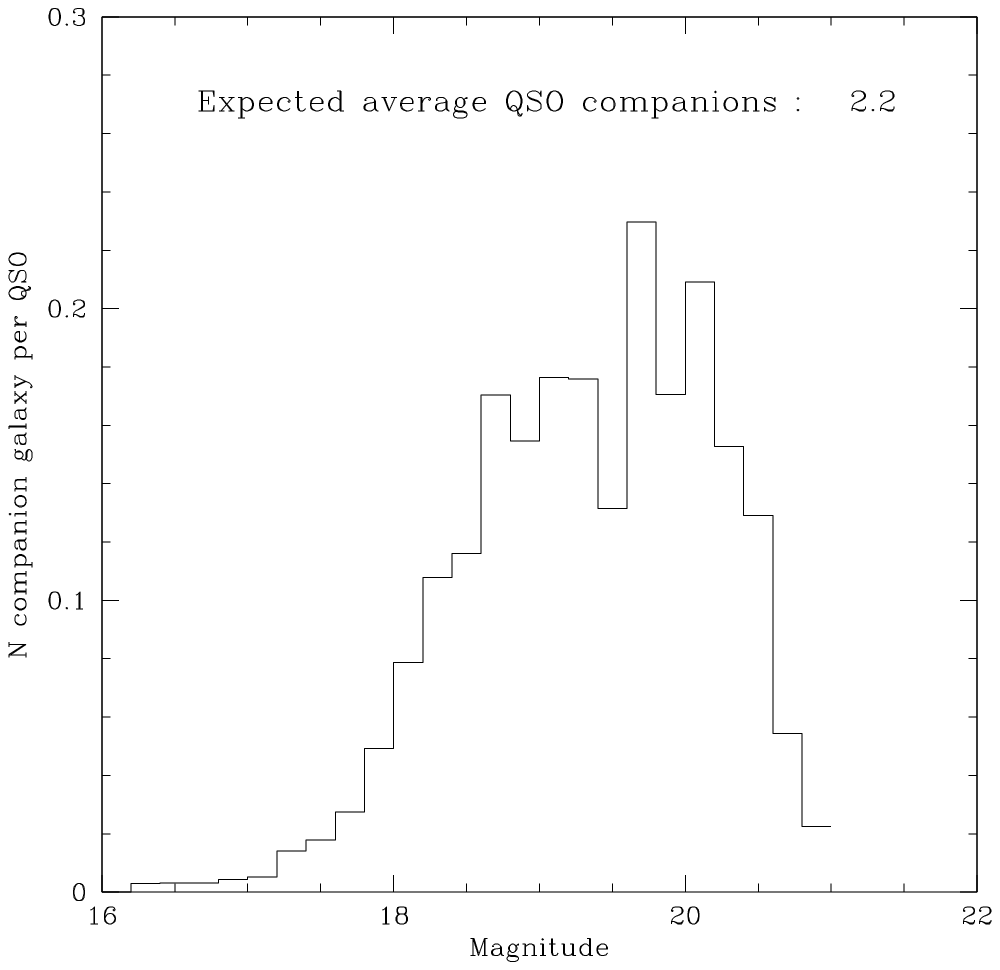}
\caption{ The expected number of associated companion galaxies for the 3112 QSO fields as a function of the galaxy magnitude. This is derived from the observed fraction of companion galaxies with spectroscopic observations and the magnitude distribution of all galaxies in the defined QSO fields (see text). 
On average we found that in a QSO  we expect to have 2.2 associated galaxies. }
\label{fig:stat_comp}
\end{figure}

We assume the observed fraction of associated companion galaxies is representative of all galaxies in the QSO fields since the selection of 
the  galaxies with a spectrum is independent on the QSO. 
Since the above fraction strongly depends on the magnitude we estimate the number of expected companion galaxies in each magnitude bin  by multiplying the above fraction by the number of all galaxies in all the 3112 QSO fields.
The distribution of expected companion galaxies as a function of magnitude is shown in Figure \ref{fig:stat_comp}.
On average we found that in a QSO of this sample we expect to find 2.2 associated galaxies.

\subsection{Emission line properties of companion galaxies}

We searched for the presence of [OII] 3727 \AA  \ and [OIII] 5007 \AA \ emission line in the spectra of companion galaxies and compared them with the galaxies in the fields that are not associated with the QSO and are in the same redshift and luminosity range as the companion galaxies.
 
\begin{figure}
\includegraphics[width=9cm]{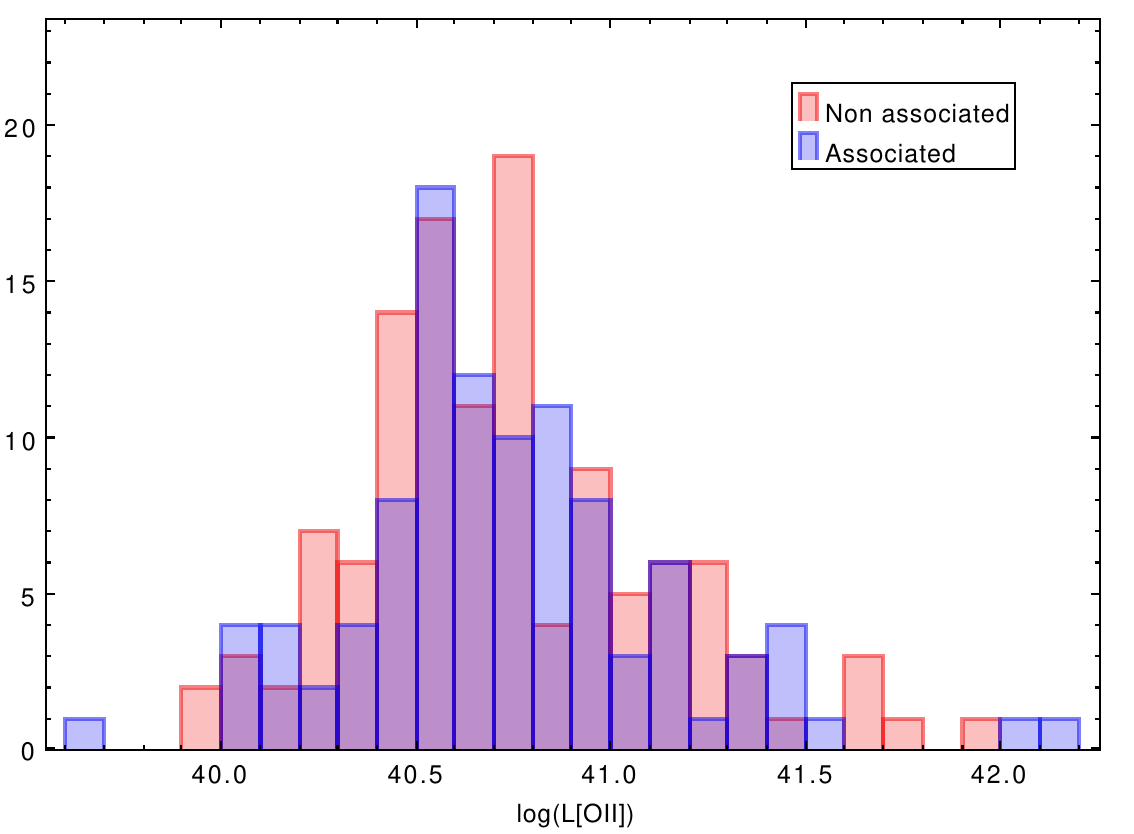}
\caption{ Distribution of [OII] luminosity of associated companion galaxies (blue) to QSO compared to that of not associated galaxies (red).}
\label{fig_lumOII}
\end{figure}

\begin{figure}
\includegraphics[width=9cm]{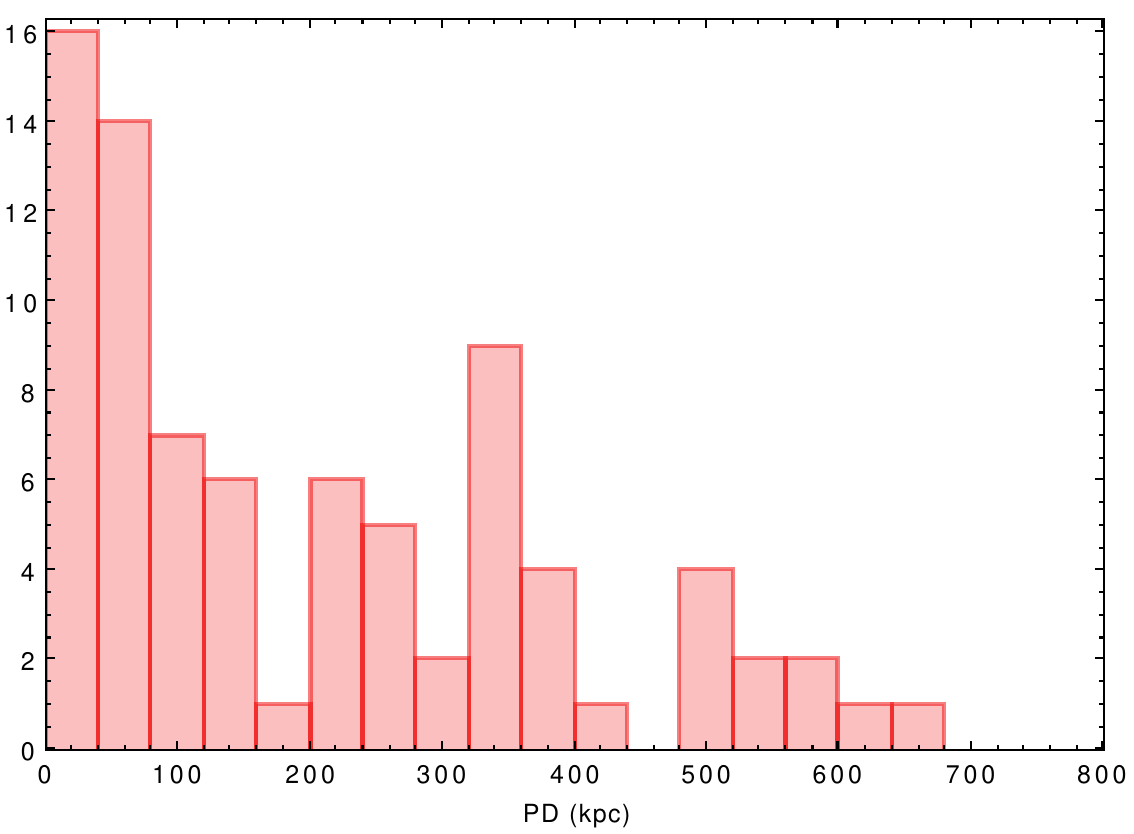}
\caption{ Distribution of Projected Distance (PD) of QSO companion galaxies showing [OII] emission lines}
\label{fig_PDOII}
\end{figure}

We limited our search to the emission lines that have a good S/N thus excluding all lines for which their EW have an error greater than 30\%.
Under these conditions, we found that the 13\% (102 objects) of companion galaxies of QSO have [OII] in emission compared to 16\% (120 galaxies) of the galaxies that are not associated. For both samples [OII]  luminosity is in the range 40~$<log(L[OII])<$~41.5 with an almost identical median value ( $log(L[OII])$~=~40.69) for the companion galaxies and ( $log(L[OII])$~=~40.68) for the non-associated objects. In Fig.~\ref{fig_lumOII}, we show the distribution of $log(L[OII])$ luminosities for both associated and non-associated galaxies. Note also that about 50\% of the companion galaxies with [OII] emission are at projected distance less than 250~kpc  see Fig.~\ref{fig_PDOII}).

A similar situation is also apparent for the [OIII] 5007 emission line. We found that the 6\% (48 objects) of companion galaxies of QSO have [OIII] in emission compared to $\sim$7\% (51 galaxies) not associated galaxies. 
Even in this case a similar range of [OIII] luminosity 
(40~$<log(L[OIII])<$~42) is found for both samples. Note that the median value of $M_r$ for our companion galaxies is -20.9, very close to the $M^*$ value for field galaxies
$M^*$~=~-20.44 \citep[e.g.][]{Blanton_2003} but lower than the galaxies that are in clusters $M^*$~=~-21.4 \citep[e.g.][]{Christlein_2003}.


It turns out therefore that there are not significant episodes of recent star formation in the companion galaxies of QSO as the same level of activity appears also in the galaxies in the fields and that have similar luminosity to the companion galaxies but are not associated to the QSO ($\Delta V >$~1000~km/s).

\section{Summary and  Conclusions}

We searched for companion galaxy candidates close to low redshift (0.1~$<z<$~0.35) QSO in the SDSS DR16 database. The association between the QSO and the galaxy is based  on having a spectrum and a projected distance less than 700~kpc from the QSO and having a radial velocity difference $\Delta V <$~1000~km/s between QSO and galaxy.
Under these conditions, and discarding a number of objects with poor quality spectra, we found 651 companions for 447 QSO fields. For the majority of the QSO only one companion is present, but in few cases the QSO is part of a relatively rich group of galaxies (see also the Appendix~A ).
Based on the  availability of spectroscopic galaxy  spectral data of the large QSO sample, on average, we expect to find $\sim$2 companion galaxies associated to the QSO. 
In order to test the fraction of possible spurious association (QSO-companion galaxy), we performed a test by mixing randomly the redshift of the companions and then search again for the associated galaxies. We run this test 1000 times and found in all cases that the number of chance projected pairs is always less than 5\%.

The average absolute magnitude of the companion galaxies is M$_r$~=~-21.8 (median $M_r$~=~-20.9) and spans from low luminosity galaxies (M$_r\sim$ -19) to massive objects (M$_r\sim$ -24). The projected distance of the companion galaxies does not appear to be correlated with either the number of companions or the difference of radial velocity with the QSO.

The majority of the companion galaxies do not exhibit spectral signature of recent star formation. Moreover, the small fraction ($\sim$ 16\%) of companion objects that have [OII] 3727 \AA ~emission line appear similar in quantity and also in terms of line luminosity to the galaxies in the field of QSO that are not associated to the target. However, the companion galaxies that exhibit emission lines are more often found at small projected distance from the QSO.  For instance while at PD~$<$~100~kpc the fraction of galaxies with emission lines is 36\% in the region 100$<$PD$<$700 kpc the fraction of companion galaxies  with emission lines is only 8\%.

From the galaxy environment spectroscopic study of these low redshift QSO, we do not find a strong link between the QSO nuclear activity and the presence of these close companion galaxies and for enhanced signature of recent star formation. Nevertheless we found that although, on average, QSO are found in poor environments, a small fraction of QSO are in relatively rich group of galaxies (see details in the Appendix~A).

\appendix
\section{QSO in rich galaxy environments}


From the selected dataset of 342 QSO with at least one companion galaxy (with PD $<$ 700 kpc and $\Delta$ V $<$  1000 km/s we found two QSO that are in a relatively rich group of galaxies. In one case (QSO SDSS J133718.48+452335.1 ) there are 12 galaxies in the field that satisfy the above conditions. In another case (QSO SDSS J092108.61+453857.3) there are 8 companion galaxies. In 3 other QSOs we found 7 companion galaxies and in other 3 cases 6 galaxies.
A summary of all these cases is given in Table \ref{tab_comp7_12}. In the following, we 
illustrate the properties for the two richer cases (QSO SDSS J133718.48+452335.1 at z = 0.317 and QSO SDSS J092108.61+453857.3 and Z = 0.175).

\begin{table}
\caption{Low Z QSO with $N_{comp}\geq 6$} 
\begin{tabular}{|l|l|r|l|r|}
\hline
  \multicolumn{1}{|c|}{SDSS} &
  \multicolumn{1}{c|}{z} &
  \multicolumn{1}{c|}{M$_r$} &
  \multicolumn{1}{c|}{Radio} &
  \multicolumn{1}{c|}{N$_c$} \\
\hline
  J133718.48+452335.1 & 0.31689 & -22.197 & RQ & 12\\
  J092108.61+453857.3 & 0.17473 & -21.516 & RL & 8\\
  J015331.78-012126.2 & 0.24274 & -21.794 & RQ & 7\\
  J222532.92+194802.6 & 0.2120 & -22.893 & RQ & 7\\
  J235941.35+120520.0 & 0.20318 & -22.139 & RQ & 7 \\
 J073623.13+392617.7 & 0.11795 & -22.453 & RQ & 6\\
  J155620.23+521520.0 & 0.2266 & -22.934 & RL & 6\\
  J232307.25+011121.0 & 0.3394 & -22.924 & RQ & 6\\
\hline\end{tabular}
\label{tab_comp7_12}  
\end{table}

\subsection{QSO SDSS J133718.48+452335.1}

\begin{figure*}[ht]
\begin{center}
\includegraphics[width=18cm]{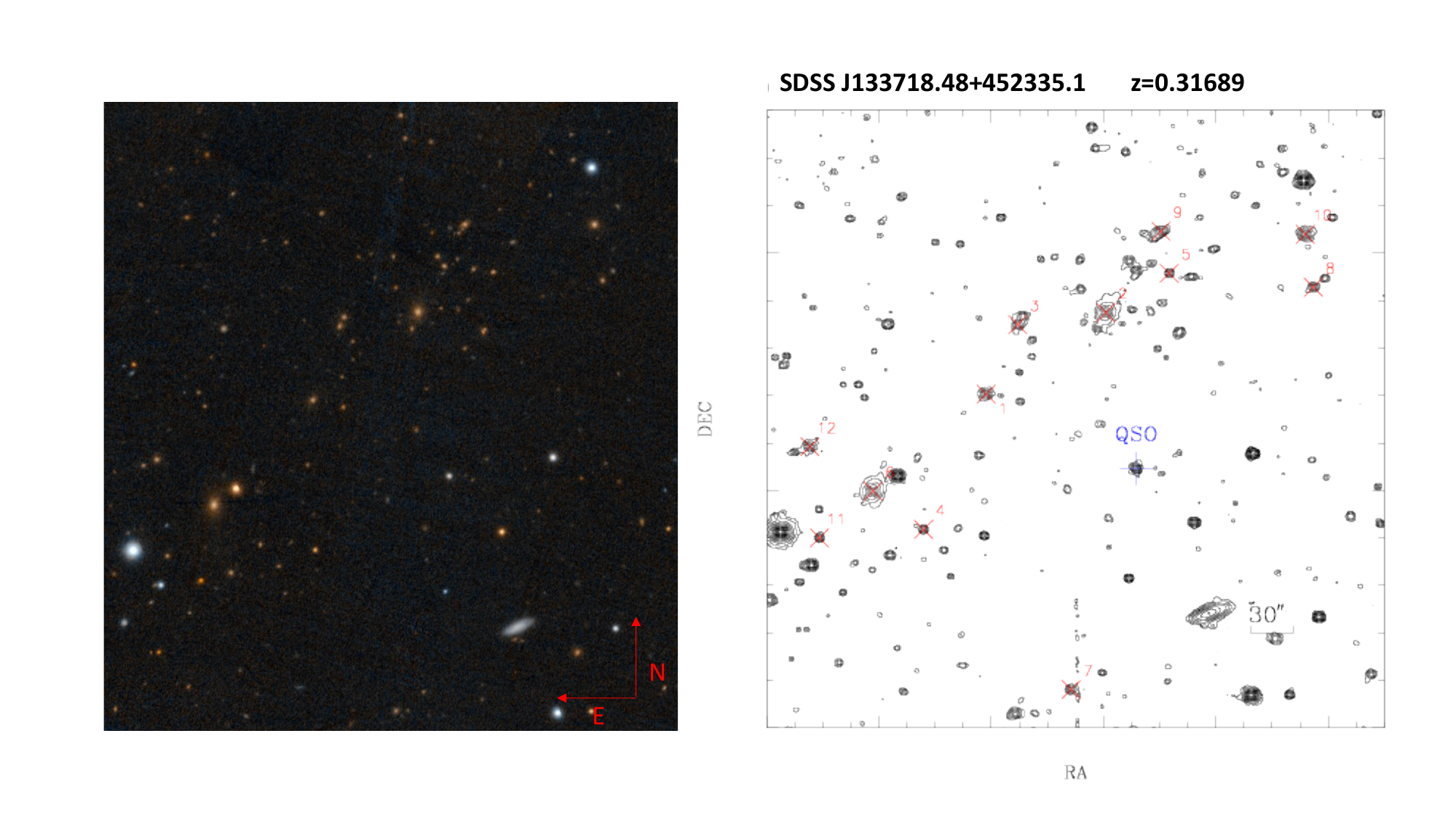}
\label{fig_qso307}
\caption { Field of the QSO  SDSS J133718.48+452335.1(FoV = 5$x$5 arcmin ; Z = 0.31689). 
There are 12 companion galaxies with PD $<$ 700 kpc and $\Delta$ V $<$  1000 km/s .
}
\end{center}
\end{figure*}

In Figure \ref{fig_qso307} we show the field (5$x$5 arcmin) around the QSO SDSS J133718.48+452335.1 ( M$_r$ = --22.2) at z = 0.31689 . There are 12 galaxies in the field with PD $<$  700 kpc and  $\Delta$ V in the range 60 to 800 km/s (median  =330 km/s). The absolute magnitude of companion galaxies are in the range --23.4 $< M_r < $ --20.7 (see Table \ref{tab_comp12}).
We search for additional possible companion galaxies in the SDSS DR16  database  in a larger field of view (PD $<$ 1.5 Mpc). There are  8 more galaxies that have $\Delta$ V $<$ 800 km/s .
The QSO appears in the central region of this rich group of galaxies. 

\begin{figure*}[ht]
\begin{center}
\includegraphics[width=18cm]{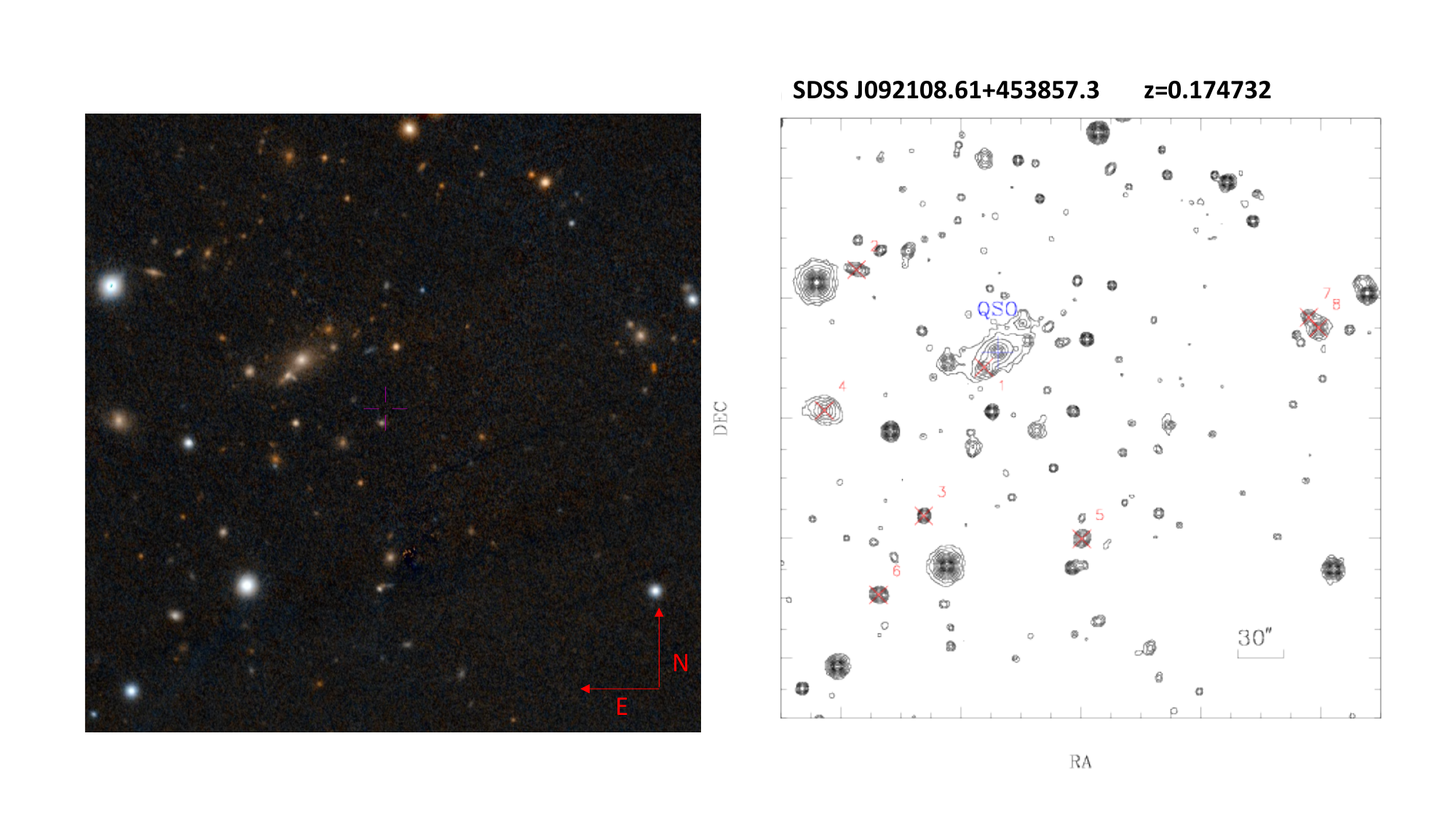}
\label{fig_qso100}
\caption { QSO SDSS J092108.61+453857.3
There are 8 companion galaxies with PD $<$ 700 kpc and $\Delta$ V $<$  1000 km/s .
Note that the companion N. 6 is itself a QSO at z= 0.1771}
\end{center}

\end{figure*}
\begin{table}[ht]
\caption{Companion galaxies of the low Z QSO SDSS J133718.48+452335.1 } 
\begin{tabular}{|l|l|r|r|r|}
\hline
  \multicolumn{1}{|c|}{Name} &
  \multicolumn{1}{c|}{z$_{comp}$} &
  \multicolumn{1}{c|}{rmag$_{comp}$} &
  \multicolumn{1}{c|}{PD} &
  \multicolumn{1}{c|}{DV} \\
 & & & Kpc~~ & km/sec \\  
\hline
  1 & 0.3171 & 19.303 & 359 & 63\\
  2 & 0.3173 & 17.823 & 385 & 123\\
  3 & 0.3141 & 19.736 & 424 & 836\\
  4 & 0.3157 & 20.156 & 461 & 356\\
  5 & 0.3182 & 20.356 & 481 & 392\\
  6 & 0.3176 & 17.635 & 546 & 212\\
  7 & 0.3160 & 19.782 & 553 & 266\\
  8 & 0.3193 & 20.208 & 575 & 722\\
  9 & 0.3183 & 18.98 & 582 & 422\\
  10 & 0.3164 & 18.63 & 667 & 146\\
  11 & 0.3153 & 20.199 & 671 & 476\\
  12 & 0.3179 & 19.496 & 676 & 302\\

\hline\end{tabular}
\label{tab_comp12}      
\centering                          
\end{table}

\subsection{QSO SDSS J092108.61+453857.3 (3C 219) and Z=0.1747 }
In Figure \ref{fig_qso100} we show the field (5$x$5 arcmin) around the radio loud QSO SDSS J092108.61+453857.3( M$_r$ = --21.5).at z = 0.17473. There are 8 objects classified as galaxies in the field with PD $<$  400 kpc and $\Delta$ V $<$ in the range 40 to 800 km/s (median  = 614 km/s). The absolute magnitude of companion galaxies are in the range --22.4 $< M_r < $ --19.8.
Note that object N. 6 (see Table \ref{tab_comp8} and Figure \ref{fig_qso100})  is  a low redshift QSO (SDSS J092113.38+453716.5) with clear broad emission lines (z=0.1771,  M$_r$ = --21.2).
This target is part of a group of galaxies studied by \citep{Madrid_06} from HST images.
Also in this case, we search for other companion galaxies at PD $<$ 1.5 Mpc. In this case no other galaxies that satisfy the association criteria are found.

\begin{table}
\caption{Companion galaxies the low Z QSO SDSS J092108.61+453857.3} 
\begin{tabular}{|l|l|r|r|r|}
\hline
  \multicolumn{1}{|c|}{Name} &
  \multicolumn{1}{c|}{z$_{comp}$} &
  \multicolumn{1}{c|}{rmag$_{comp}$} &
  \multicolumn{1}{c|}{PD} &
  \multicolumn{1}{c|}{DV} \\
 & & & Kpc~~ & km/sec \\  
\hline
  1 & 0.1731 & 17.872 & 25 & 489\\
  2 & 0.1774 & 18.964 & 204 & 800\\
  3 & 0.1728 & 19.068 & 219 & 579\\
  4 & 0.1769 & 17.267 & 227 & 650\\
  5 & 0.1732 & 18.897 & 251 & 459\\
  6 & 0.1771 & 18.449 & 337 & 710\\
  7 & 0.1746 & 19.825 & 386 & 39\\
  8 & 0.172 & 18.052 & 392 & 819\\
  
  \hline\end{tabular}
\label{tab_comp8}      
\centering                          
\end{table}

\bigskip

We also searched for other galaxies at PD $<$ 1.5 Mpc in the other 6 cases with 6 and 7 companion
galaxies (see Table \ref{tab_comp7_12}). For 3 QSO fields, we found a number of additional companion galaxies: 16 new members for SDSS J015331.78-012126.2, 9 members for SDSS J235941.35+120520.0  and 24 members for SDSS J073623.13+392617.7.
In the remaining three cases no other companion galaxies are reveled.

%

\begin{acknowledgements}

We thank the anonymous referee for very helpful comments and suggestions.
Funding for the Sloan Digital Sky Survey IV has been provided by the Alfred P. Sloan Foundation, the U.S. Department of Energy Office of Science, and the Participating Institutions. SDSS-IV acknowledges
support and resources from the Center for High-Performance Computing at 
the University of Utah. The SDSS web site is www.sdss.org.

SDSS-IV is managed by the Astrophysical Research Consortium for the 
Participating Institutions of the SDSS Collaboration including the Brazilian Participation Group, the Carnegie Institution for Science, 
Carnegie Mellon University, the Chilean Participation Group, the French Participation Group, Harvard-Smithsonian Center for Astrophysics, 
Instituto de Astrof\'isica de Canarias, The Johns Hopkins University, Kavli Institute for the Physics and Mathematics of the Universe (IPMU) / 
University of Tokyo, the Korean Participation Group, Lawrence Berkeley National Laboratory, 
Leibniz Institut f\"ur Astrophysik Potsdam (AIP),  
Max-Planck-Institut f\"ur Astronomie (MPIA Heidelberg), 
Max-Planck-Institut f\"ur Astrophysik (MPA Garching), 
Max-Planck-Institut f\"ur Extraterrestrische Physik (MPE), 
National Astronomical Observatories of China, New Mexico State University, 
New York University, University of Notre Dame, 
Observat\'ario Nacional / MCTI, The Ohio State University, 
Pennsylvania State University, Shanghai Astronomical Observatory, 
United Kingdom Participation Group,
Universidad Nacional Aut\'onoma de M\'exico, University of Arizona, 
University of Colorado Boulder, University of Oxford, University of Portsmouth, 
University of Utah, University of Virginia, University of Washington, University of Wisconsin, 
Vanderbilt University and Yale University.
\end{acknowledgements}

\bibliographystyle{aa}
\bibliography{references.bib}

\end{document}